# Structure of a Zn monolayer on Ag(111) and Ag(110) substrates: an AES, LEED and STM study

Maradj Hicham[1,2], Carole Fauquet[2], Mostefa Ghamnia[1], Benedicte Ealet[2], Nabil Rochdi[3], Sebastien Vizzini[4], Jean-Paul Biberian[2], Bernard Aufray[2] and Haik Jamgotchian[2*]

[1] Laboratoire LSMC, Université d'Oran es-sénia, 31100, Oran, Algeria

[2] Aix-Marseille Université, CNRS, CINaM UMR 7325, 13288 Marseille, France

[3] SIAM, Département de Physique, Faculté des Sciences Semlalia, Université Cadi Ayyad, Marrakech – Morocco

[4] Aix-Marseille Université, CNRS, IN2MP UMR 7334, Marseille, France

* Corresponding auteur: jamgotchian@cinam.univ-mrs.fr

**Abstract**

Auger Electron Spectroscopy, Low Energy Electron Diffraction and Scanning Tunneling Microscopy have been used to study the atomic structure of a Zn monolayer deposited on Ag(111) and Ag(110) substrates at room temperature. On both faces, there is formation of a close packed monolayer of Zn covering the entire substrate surface and giving rise to specific Moiré patterns. From a comprehensive LEED and STM data analysis, we deduce that the Zn monolayer adopts a (111) structure equivalent to a pure Zn layer rotated with respect to the silver substrate, of about 1.5° on the Ag(111) face and of about 4.5° on the Ag(110) face giving rise respectively to ($\sqrt{156}\times\sqrt{156}$)R18° and c(12x6) superstructures.

## Introduction

Due to their important potential applications in the domain of catalysis or magnetism, the deposition of metal thin films on metallic substrates is of great interest. It is also very important from a fundamental point of view to the understanding of epitaxy mechanisms. Since the discovery of graphene [1,2], and equivalent silicene [3], germanene [4], [5], [6] …, films of one monolayer thickness have been studied with a particular attention due to their potential unexpected and new electronic properties: high 2D conductivity, Dirac fermions, Mott insulators [7], [8,9] ...

Generally the deposition of a metal A on a metallic substrate B is related to the corresponding relative segregation phenomenon of both elements in AB alloy in particular when the substrate temperature allows atomic exchanges close to the surface. In this framework there are three driving forces which allow to predict the element which will have a tendency to segregate at the surface [10,11] (i) the surface energy difference; (ii) the atomic radius difference and (iii) the order or phase separation tendency. It is the coupling between these three effects which drives (in most cases) the behavior of the system. Many experimental studies have been interpreted at the light of these three driving forces and predictable maps have been drawn up for many bimetallic systems [12].

In the special case of the deposition of one monolayer of an element having a surface energy lower than the one of the substrate, the atomic structure of the deposited monolayer is strongly dependent on the atomic radius difference between elements. Indeed when the deposited element is larger than the one of the substrate (which is the most common case) the monolayer forms Moiré patterns directly related to the difference between adsorbate and substrate atomic radii (due to the Vernier effect). In contrary, when the deposited element is smaller than the one of the substrate (few systems present this characteristic), there is a coherent epitaxy (pseudomorphy) of the monolayer with formation of periodic linear defects in relation to the strain relaxation [13]. In other words the monolayer of the deposited element presents in this case the same atomic parameters than the one of the substrate. In this paper we experimentally show that it is not always the case since we do not observe the expected coherent epitaxy when a Zn monolayer is deposited at room temperature on Ag(111) et Ag(110). The Zn on Ag system corresponds to the case where the deposited element presents a lower surface energy than the substrate one (0.99 and 1.25 J.m$^{-2}$ respectively [14]) and Zn atoms are smaller than Ag atoms (0.266 and 0.289 nm in diameter respectively). Instead we observed Moiré patterns as expected when adsorbed atoms are larger than substrate ones.

In 1994, Zn deposition on Ag substrates was first studied by Kourouklis et al. [15] using Auger Electron spectroscopy (AES), Low Energy Electron Diffraction (LEED) and Temperature Programmed Desorption (TDS). The main results can be summarized as follows: on the three low index Ag faces (110), (100) and (111) the growth kinetics (i.e. the variation of Ag and Zn intensities Auger peaks versus the deposition time) show at the beginning of the growth, a surprising delay; i.e. in the onset of the growth, there is no detected Zn Auger signal, whereas the Ag signal starts to decrease with respect to its initial value. No superstructure was observed by LEED at all deposited Zn amounts, substrate temperatures and Ag surface orientations. From their results the authors propose a growth mode starting with 3D Zn clusters formation on Ag surfaces.

Here we present a set of new experimental data obtained by AES, LEED and Scanning Tunneling Microscopy (STM) of Zn deposition at room temperature (RT) on Ag(111) and Ag(110) substrates. This paper is mainly dedicated to the atomic structure of the Zn monolayer on both silver faces, which was not observed by Kourouklis et al. [15]. In a forthcoming paper we will focus on the first stages of the Zn growth.

In the following, we first describe the experimental set-up and then present and discuss the experimental results obtained by AES, LEED and STM after the deposition of one Zn monolayer on Ag(111) and Ag(110).

**<u>Experimental set-up</u>**

The experiments are performed in three different ultra-high vacuum (UHV) chambers connected all together (P < $10^{-9}$ mBar). The main chamber contains the Zn evaporator, an Auger spectrometer (CMA) and a LEED optics. Attached to this main chamber there is a STM (Omicron STM-1) working at room temperature with a tungsten tip cleaned by heating. The sample surface cleaning is performed in a third chamber also attached to the main chamber. The sample preparation consists of repeating surface Ar ion sputtering (1 hour, 550eV) followed by annealing at 800K for 1h in UHV conditions. The surface cleanness is checked by AES and LEED. The sample is heated via a tungsten wire placed under the sample holder and the sample temperature is controlled by a thermocouple located close to the sample. The evaporation of Zn is performed from an alumina crucible heated with a tungsten filament. The evaporation rate and the calibration of Zn amounts are deduced from Auger, LEED and STM measurements (see below). The experimental protocol to record growth kinetics consists of maintaining the sample alternatively in front of the Zn evaporator for a constant time (2 min) then in front of the Auger analyzer to record the chemical surface evolution and finally in front of the LEED optics to observe the surface structure evolution. All the STM images are recorded in the filled state mode.

## **Zn deposition on Ag(111)**

### *The AES-LEED study*

Figure 1 shows the variations at room temperature, of the peak-to-peak Auger signal intensities of Ag (355 eV), and Zn (56eV) versus Zn deposition time. The shape of this Auger growth kinetics is very close to the one presented by Kourouklis et al. [15] on the Ag(100) face at room temperature. The intensity of the Zn (56eV) Auger peak presents at the beginning of the growth the same delay. After this delay, there is a quasi linear increase of the Zn intensity up to a slope change (a break) at about 10 min of deposition. After this break the Zn intensity increases slowly up to plateau. In a concomitant way, the intensity of the Ag Auger peak is continuously decreasing down a value close to zero. From these curves and from LEED and STM characterizations (see below) we deduced that the Zn ML is obtained after about 10 min of Zn deposition time (pointed out by a black vertical line on Figure 1).

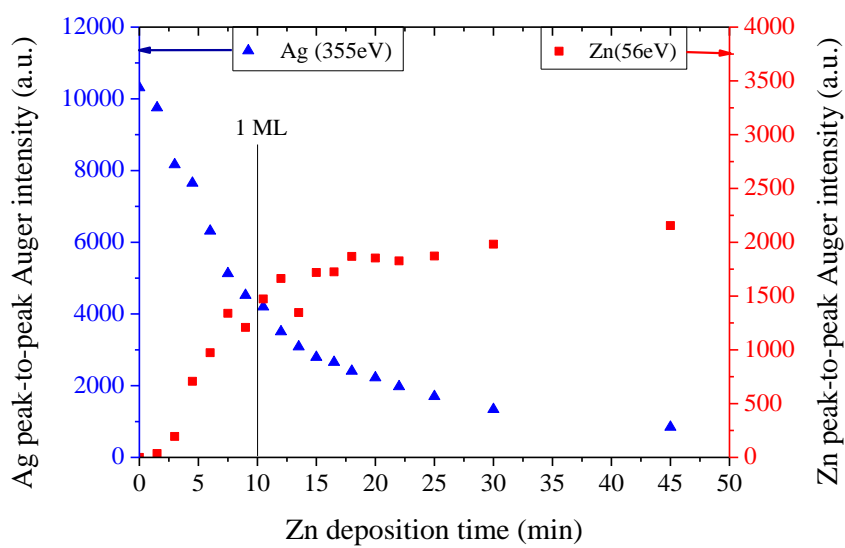

*Figure 1: Auger Signal intensities as a function of the Zn deposition time at room temperature*

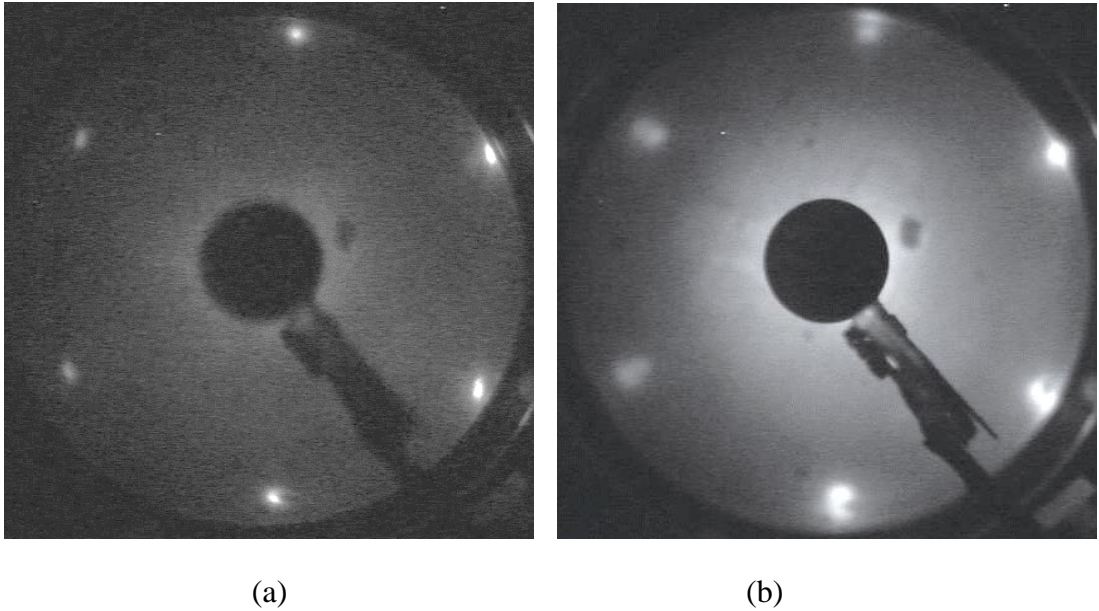

(a)                  (b)

*Figure 2:* LEED patterns of an Ag(111) surface a) recorded just after the cleaning procedure (Ep= 47 eV) b) recorded after 10 min of Zn deposition (Ep= 43 eV)

LEED controls were carried out after each Auger analysis. Figures 2a and 2b show the LEED patterns recorded just after the cleaning procedure of the Ag(111) sample and after 10 min of Zn deposition. The extra-spots start to appear after 6 min of deposition going through a maximum (Figure 2b) to completely disappear with an increase of the background after 13 min of deposition. At first glance this LEED pattern is in line with a dense (111) Zn plane in epitaxy on the Ag(111) substrate. Indeed, the ratio between the lattice parameters deduced from the LEED pattern of Ag and Zn is about 1.1 in agreement with the expected ratio between Ag (0.289 nm) and Zn (0.266 nm) interatomic distances (1.086). Nevertheless, a slight elongation of Zn spots indicates small rotations of the Zn layer with respect to the substrate.

### *The STM study*

The STM study has been performed when the LEED pattern shows the superstructure reported on Figure 2b, i.e. after 10 min of deposition time.

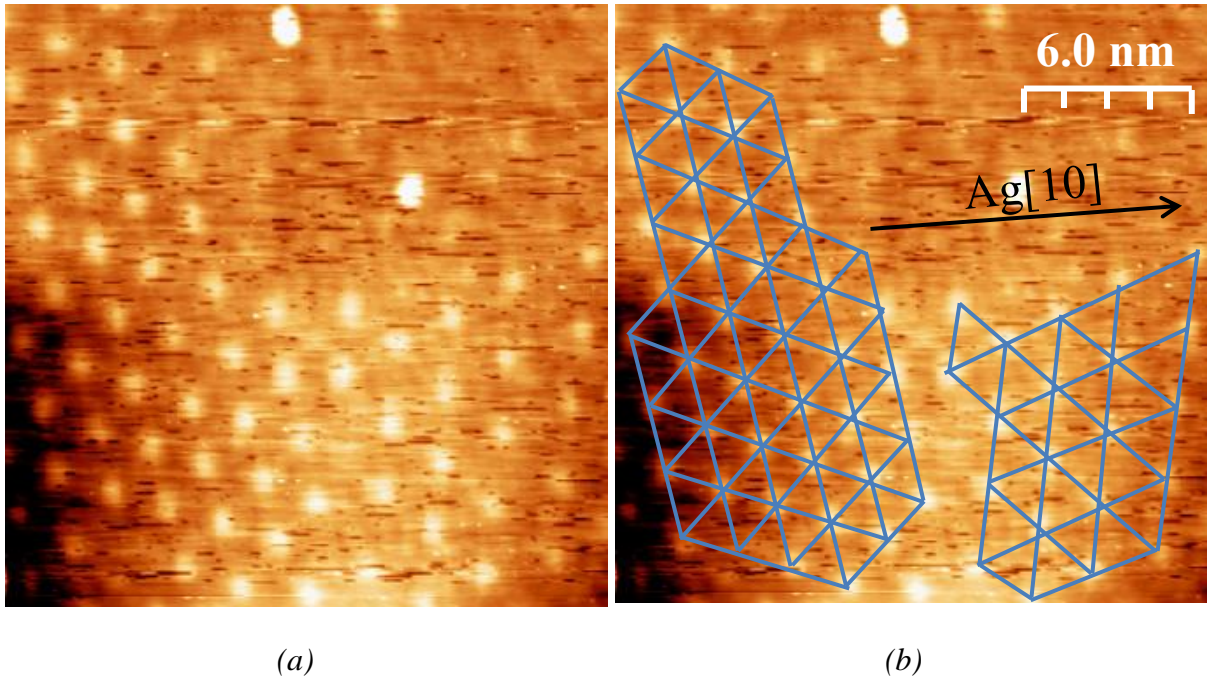

*(a)* *(b)*

***Figure 3:*** *STM image of the Ag(111) surface recorded after 10 min of Zn deposition time*

Figure 3a displays a filled states STM image recorded after 10 min of Zn deposition time. This image shows two domains of Moiré Patterns (MP) highlighted by blue grids on Figure 3b. The orientation of the two domains are symmetric with respect to the [10] direction of silver (pointed out on Figure 3b by a black arrow).

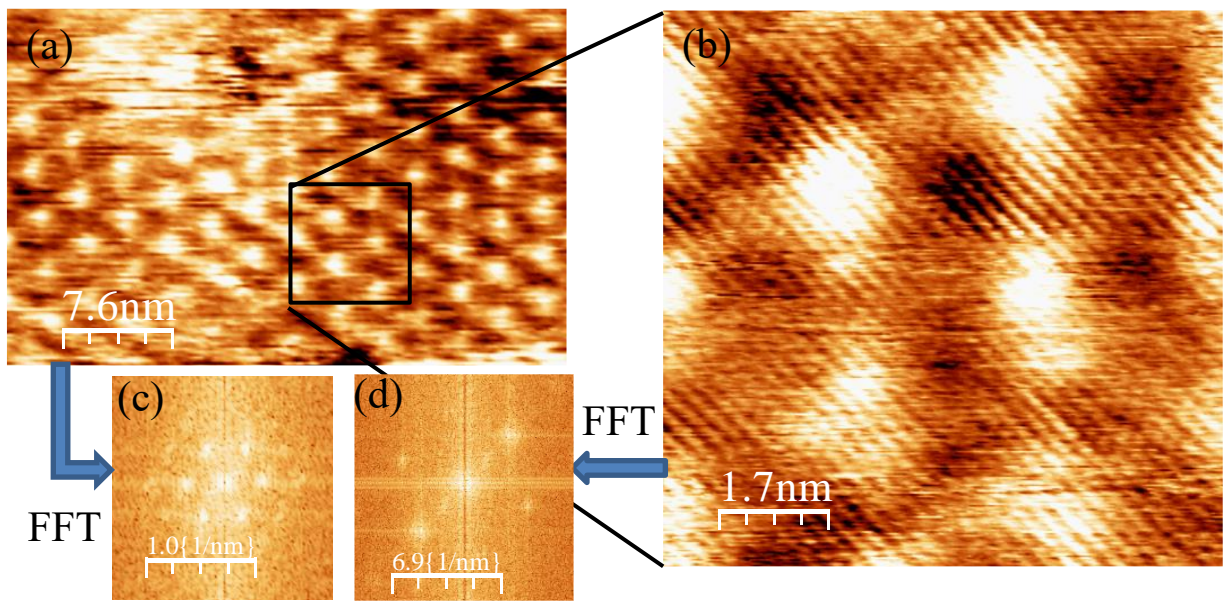

***Figure 4:*** *STM images a) at low resolution showing the MP and b) with atomic resolution. Their corresponding FFT are shown on (c) and (d)*

Figure 4a is a low resolution STM image of the MP and Figure 4b is a high resolution (HR) STM image of a part of the same area. The corresponding FFT of both images are shown on Figure 4c and 4d respectively. From Figure 4c, we deduce the characteristics of the MP: the average distance between two white protrusions ($\lambda \sim 3.4$ nm) and its angle relative to the Ag substrate ($\gamma \sim 22°$). From Figure 4d, we deduce the characteristics of the Zn monolayer: the interatomic distance of atoms ($d_{Zn} \sim 0.26$ nm, close to Zn interatomic distance 0.266 nm) and its angle of rotation relatively to the Ag substrate ($\alpha \sim 1.5°$). A detailed observation of this image reveals a dispersion of the MP in size and in orientation. The FFT used here give an average value of these parameters.

In order to determine the MP dispersion we have measured its size ($\lambda$) and angle ($\gamma$) directly on eleven different areas on the sample. The corresponding values are reported in Table 1. We deduced an average angle ($\gamma$) of the MP relatively to the Ag [10] direction ($\gamma=18° \pm 7°$) and an average size of the unit cell ($\lambda= 3.3$ nm $\pm 0.8$nm).

| # | 1 | 2 | 3 | 4 | 5 | 6 | 7 | 8 | 9 | 10 | 11 | average | dispersion |
|---|---|---|---|---|---|---|---|---|---|---|---|---|---|
| λ (nm) measured | 3.9 | 2.7 | 3.3 | 2.6 | 3.6 | 3.1 | 3.3 | 3.9 | 3.9 | 2.5 | 3.5 | 3.3 | ± 0.8 |
| γ (°) measured | 15 | 13 | 22 | 20 | 14 | 20 | 16 | 19 | 14 | 23 | 25 | 18 | ± 7 |
| α (°) calculated | 1.1 | 1.3 | 1.8 | 1.9 | 1.0 | 1.7 | 1.2 | 1.3 | 1.0 | 2.3 | 1.9 | 1.5 | ± 0.8 |
| $d_{Zn}$ (pm) calculated | 270 | 261 | 267 | 261 | 268 | 265 | 266 | 270 | 270 | 261 | 269 | 266 | ± 5 |

*Table 1: MP parameters measured on STM images recorded on eleven different areas. γ is the rotation angle of the MP with respect to the Ag[10] direction and λ is the Moiré size. α is the rotation angle of the Zn monolayer with respect to the Ag[10] direction and $d_{Zn}$ is the distance between two Zn atoms (α and $d_{Zn}$ are deduced from the experiments using relations 1 and 2 see below)*

*Model*

K. Herman [16] studied the theoretical geometric relations of an MP formed by the superposition of two periodic structures. He deduced two relations linking the parameters of the two periodic structures and of the Moiré one. When applied to two triangular structures like Ag(111) and Zn(111) the relations become:

$$\tan \alpha = \frac{\sin \gamma}{(\lambda/d_{Ag} + \cos \gamma)} \quad (1)$$

$$d_{Zn} = \frac{\lambda}{\sqrt{1+(\frac{\lambda}{d_{Ag}})^2 + 2(\frac{\lambda}{d_{Ag}})\cos \gamma}} \quad (2)$$

For each of the eleven different values, the relations allow to calculate α (the rotation angle of the Zn monolayer with respect to the substrate) and $d_{Zn}$ (distance between two Zn atoms) which both are reported on Table 1. Calculated α values are between 1.0° to 2.3° and

$d_{Zn}$ from 0.261 nm to 0.270 nm. It is noteworthy that the observed dispersion is larger than the measurement precision showing that the unit cell of the Zn monolayer fluctuates in size and angle (more in angle than in size).

Figure 5a and 5b display the theoretical variations of $\gamma$ and $\lambda$ as a function of $\alpha$ in the case of the two extremum values of $d_{Zn}$ (the blue and the red lines) and for its average value (the black line). The eleven experimental values of the MP are also reported. These curves indicate that the average value $d_{Zn}$ corresponds to its bulk value in the hexagonal (0.01) plane. It shows too, that a variation of few degrees of $\alpha$ gives rise to a large variation of $\gamma$ and $\lambda$. Moreover, the $d_{Zn}$ variation remains weak with respect to $\alpha$ variation. As a consequence, the dispersion of $\alpha$ and $d_{Zn}$, explains the elongated shape of the Zn spots of the LEED pattern.

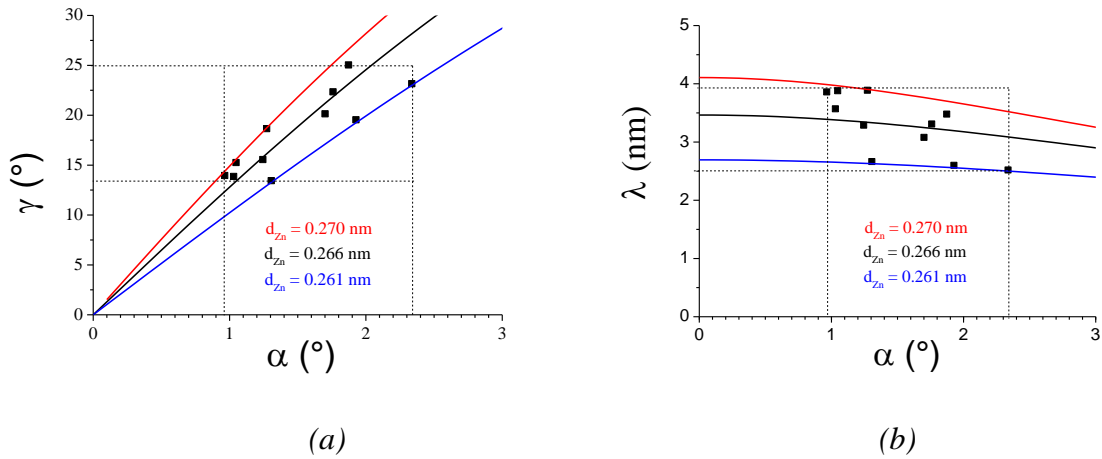

*(a)*        *(b)*

*Figure 5: Theoretical variations of the MP angle (a) and of the Moiré periodicity (b) versus the angle of rotation of the Zn monolayer with respect to the Ag. The black points correspond to the experimental values measured on different STM images*

Figure 6 is a ball model of a Zn close packed monolayer on an Ag(111) plane in the special case of the average parameters: $\alpha = 1.5°$ and $d_{Zn} = 0.266$ nm (Table 1). This superstructure corresponds to a $(\sqrt{156} \times \sqrt{156})R18°$ in the Wood's surface structure notation. Both the rotation angle (18°) and the size (($\sqrt{156}\, d_{Ag} = 3.6$ nm) of this simulated MP are in very good agreement with the average values determined experimentally from different STM images.

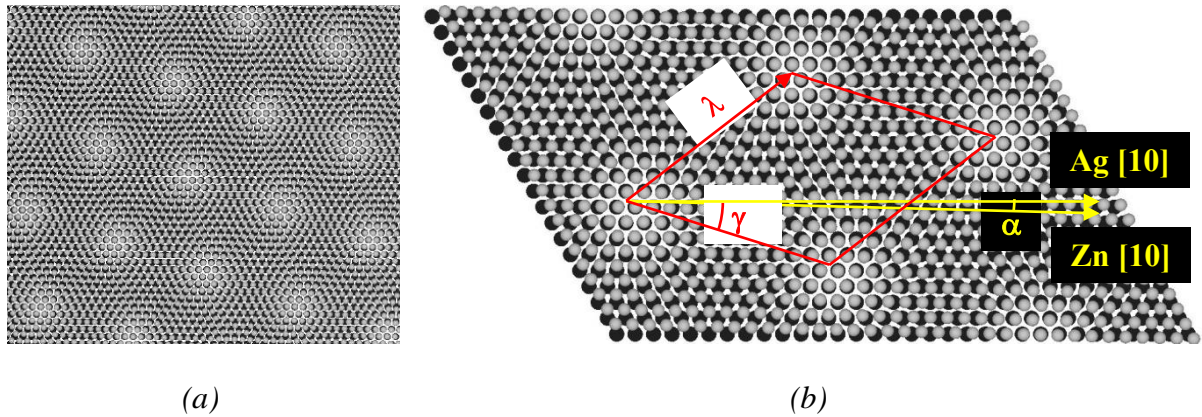

*(a)*          *(b)*

***Figure 6****: Ball model of a dense plane of Zn (grey balls) rotated 1.5° relatively to silver (black balls). The unit cell of the Moiré pattern forming the (( √156x √156)R18° superstructure is highlighted in red. a) Large view b) Enlarged view showing details of the epitaxy. In red: the Moiré unit cell. In yellow: the Zn and the Ag [10] directions*

Usually a superstructure with two domains gives two symmetrical spots on the LEED pattern. We did not observe this expected double spots. This could be due to the small angle of rotation of the Zn plane and to its quite large dispersion. Let us remark that just from the LEED diffraction pattern we expected a Moiré pattern without the rotation with a size corresponding to a Vernier effect: 13 atoms of Zn correspond to 12 silver atoms giving rise to a Moiré size ~3.4 nm (13/12=1.083 close to 0.289/0.26=1.086) which is close to the experimental one (3.2 nm).

### **Zn deposition on Ag(110)**

#### *The AES-LEED study*

Figure 7 shows the variations at room temperature of the peak-to-peak Auger signal intensities of silver (355 eV) and Zn (56 eV) versus Zn deposition time. As for the (111) orientation we observe a similar shape of the growth kinetics i.e. first a delay at the beginning of the growth, then a quasi linear increase up to a slope change (at about 10 min) and finally a plateau after 13 min of Zn deposition. From these curves and from LEED and STM characterizations (see below) we deduced that the Zn ML is obtained after about 10 min of Zn deposition time pointed out by a black vertical line on Figure 7.

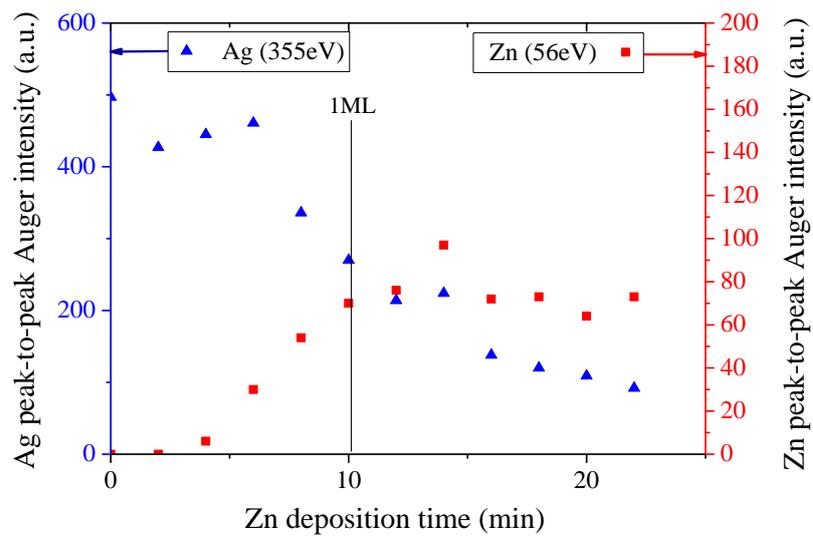

*Figure 7: Auger signal intensities of Zn and Ag as a function of the Zn deposition time recorded at room temperature*

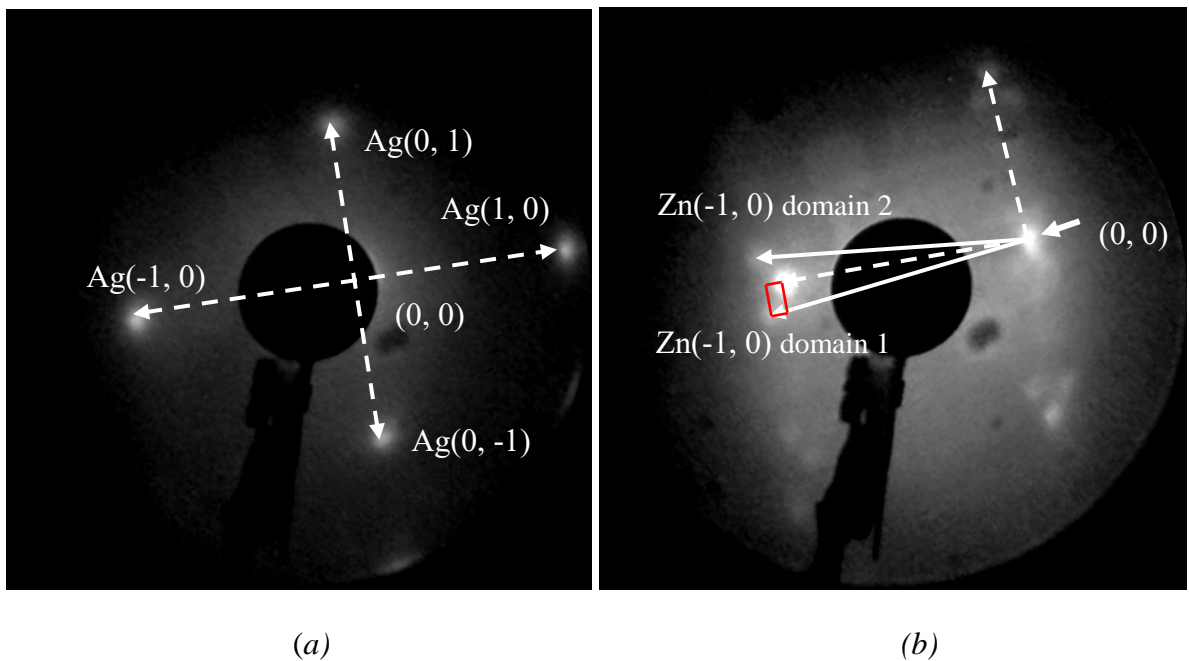

(*a*)                  (*b*)

*Figure 8: LEED patterns observed a) before Zn deposition (Ep= 50 eV) b) after 10 min of Zn deposition time on Ag (110) (Ep = 37 eV). The c(12x6) unit cell is highlighted in red*

LEED controls were carried out after each Auger analysis. Figure 8a shows the LEED pattern of Ag(110) surface after the cleaning procedure i.e. before Zn deposition. Figure 8b is the LEED pattern observed after 10 min of Zn deposition time. In comparison with Figure 8a the sample holder has been rotated a few degrees in order to show the extra spots around the (0 0) Ag spot. The extra spots observed on this LEED pattern start to appear after 3 min of deposition time, until sharp spots appear after 10 min. The LEED pattern completely disappears with an increase of the background after 15 min of deposition. These extra spots correspond to a c(12x6) superstructure.

*The STM study*

The STM study was performed when the LEED pattern showed the superstructure reported on Figure 8b i.e. after 10 min of Zn deposition time. One observes two domains of stripes which directions are indicated by blue arrows on the STM image. Both directions are symmetrical with respect to [110] direction of Ag. The orientation of the Ag substrate is given by the atomic resolution of the Ag surface before Zn deposition, inserted in the bottom right of the image. These two domains are also visible on the LEED pattern shown in Figure 8b through the two symmetrical extra spots with respect to Ag(-10) direction. Note that the ratio of the spot distances measured on LEED for Zn and Ag is ~1.1 which corresponds to the ratio between Ag and Zn interatomic distances (0.289/0.266=1.086).

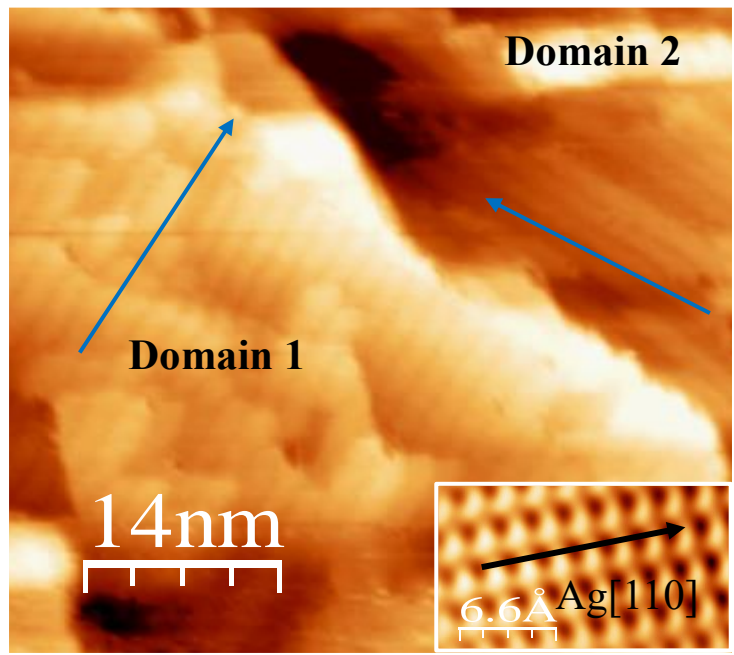

*Figure 9: STM image of the Ag(110) surface after 10 min of Zn deposition. The insert corresponds to the atomic resolution of a bare Ag(110) which shows the orientation of the substrate*

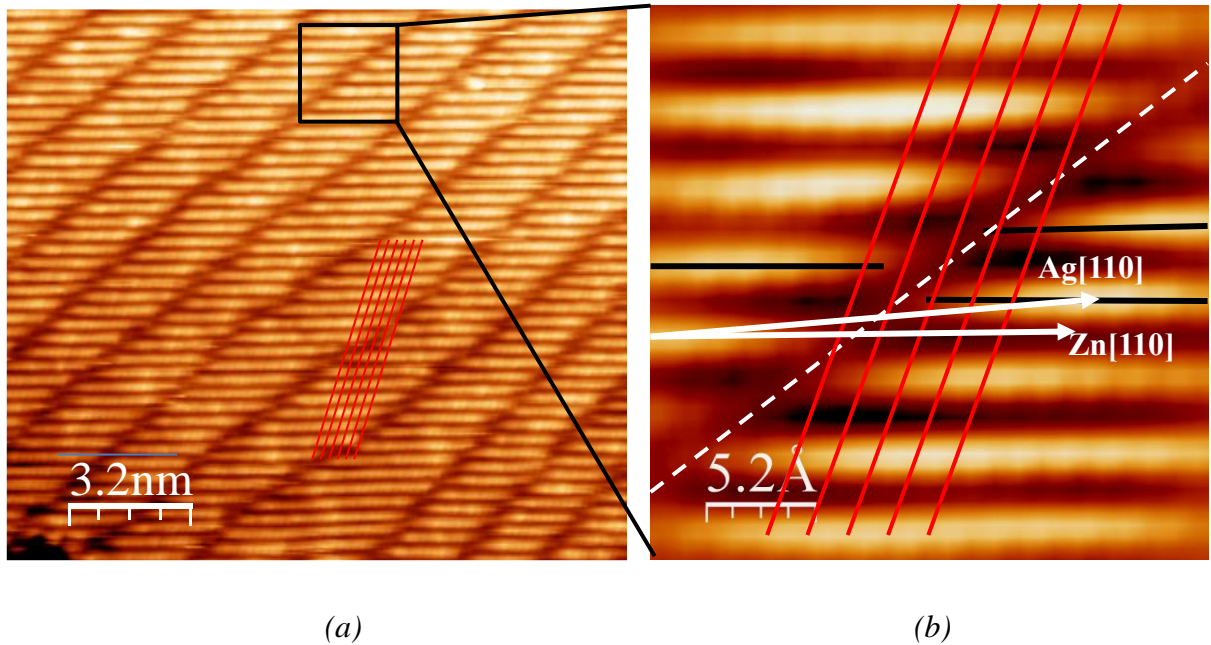

*(a)*          *(b)*

*Figure 10: a) High resolution STM image of domain 1. b) Blow-up of part of the STM image (filtered image)*

Figure 10a is a high resolution STM image of domain 1 in which we observe Moiré patterns in the shape of diagonal and horizontal lines. There is a large dispersion of the distance between diagonal lines (from 1.9 nm to 3.7 nm with an average value of 2.2 nm), while the distance between the horizontal lines is constant and close to the one between two dense silver rows (~ 0.4 nm). Figure 10b is an FFT filtered STM image of a blow-up of part of the STM image Figure 10a. This HR STM image shows that the Moiré line (highlighted by a white dashed line), is generated by a periodic shift of the horizontal lines (highlighted by black lines). The average distance between two protrusions along a horizontal line is close to $d_{Zn}$ (0.26 nm). The red lines show a perfect and regular alignment of protrusions across the diagonal Moiré lines.

*Model*

Taking into account the LEED pattern, the deposited amount of Zn and the STM images we assume that, as for Ag(111) surface, a dense triangular plane of Zn is present on the Ag(110) surface. The best agreement with experimental data is obtained by a superposition of a Zn pseudo-triangular monolayer (monoclinic cell: 0.277x0.260 nm$^2$ angle 122°) on an Ag(110) surface. A rotation of $\alpha = 3.7°$ of the Zn monolayer is the origin of the formation of Moiré lines and of a c(12x6) superstructure as shown in Figure 11. The white rectangles display the unit cell of the c(12x6) superstructure. The white diagonal dashed line drawn in the unit cell indicates the diagonal Moiré line. The red lines highlight the alignment of the Zn[11] and the Ag[11] directions. For a better understanding of the model, the top part of Figure 11b shows the Zn monolayer without underlying silver while the bottom part shows the silver top layer without the Zn monolayer. The corresponding unit cells of Zn and Ag are indicated in black.

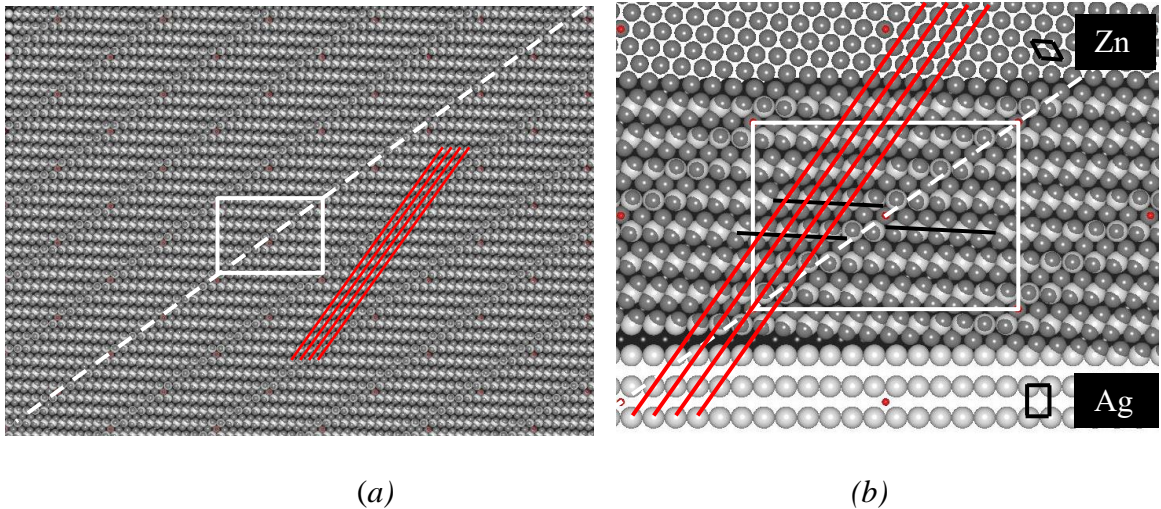

(a)                                    (b)

**Figure 11:** *Ball model of a pseudo triangular Zn monolayer (0.277 x 0.260 nm$^2$, 122°, $\alpha =$ 3.7°) deposited on the Ag(110) surface generating a c(12x6) superstructure and the associated Moiré pattern. a) Large view; b) Blow-up view showing details of the epitaxy*

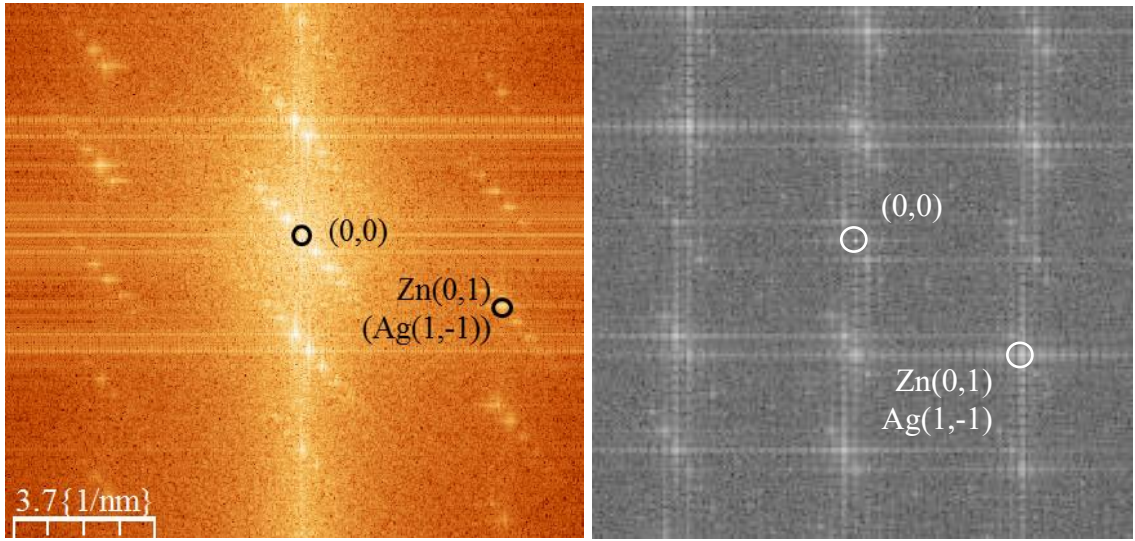

(*a*)  (*b*)

***Figure 12:*** *FFT obtained: a) from the STM image given in Figure 10a and b) from the ball model given in Figure 11a*

Figure 12 compares the FFT's of the STM image (Figure 10a) and of the ball model (Figure 11a). The good agreement between the two FFTs and the LEED pattern validates the proposed ball model.

Hermann et al. [16] did not examine from a theoretical point of view the case of Moiré patterns formed by triangular structures on rectangular ones. In this case, we have shown that there is formation of parallel strips for which width and orientation are strongly dependent of relative rotation and interatomic distances of the two structures. Thus, as on the Ag(111) surface, a slight rotation of the Zn monolayer gives rise to a large variation of the strips width and orientation with respect to the Ag[10] direction. The angular and size dispersions of the Moiré pattern observed locally show that the epitaxy on Ag(110) could be a compromise between the rotation and a weak deformation of the Zn unit cell with a possible transition from triangular to monoclinic structure due to the buckling and the assymetry of the Ag(110) surface.

On the LEED pattern, the extra spots associated with the two domains of the Zn monolayer on the Ag(110) surface are clearly distinct, unlike the Ag(111) LEED pattern extra spots, which are quasi indistingishable. This difference is due to the fact that the angle of rotation of Zn monolayer on Ag(110) is larger than on Ag(111) substrate, 3.7° v*s* 1.5°.

**General discussion**

On Ag(111) and on Ag(110) we observe the same delay at the beginning of the Auger growth kinetics that Kourouklis *et al.* did on Ag(100). They interpreted this delay by a growth mode Volmer-Weber type; *i.e.* by the formation of Zn 3D islands. At the light of our results we think that this conclusion is not correct, since we observe an epitaxy of a Zn dense (111) monolayer on Ag(111) as well as on Ag(110) faces which show the tendency of Zn atoms to wet the surface.

From a thermodynamics point of view, the formation of a quasi pure Zn monolayer is in line with a Zn surface energy much lower than the Ag one (0.99 and 1.25 J.m$^{-2}$ respectively [14]). Note that the two other driving forces (size and chemical effects), are not in favor of Zn surface segregation. Indeed, the size ($d_{Zn} < d_{Ag}$) and chemical effects (tendency to order) will not yield a Zn surface segregation.

From a structural point of view, on both faces, the MP's are not regular and present local dispersions in size and direction. This could be due to local rotations of the Zn ML and, to a lesser degree, to a size variation of the Zn unit cell. This means that on both faces, the chemical interactions between Ag substrate and Zn ML are weak: in other words, the Zn ML is not strongly linked to the Ag substrate. Nevertheless, this "floating-like" behaviour seems to be more evident on the (111) face than on the (110) one. In any case, this almost "floating-like" behaviour reveals strong in plane Zn-Zn bonds. This could be in relation with the hexagonal compact bulk structure of Zn which presents a large c/a ratio: 1.86, the highest of all metallic elements with an hexagonal compact structure (1.63).

**Conclusion**

We have studied the deposition of a Zn monolayer on Ag(111) and Ag(110) substrates by AES, LEED and STM. The growth kinetics profile obtained by AES is in agreement with previous work by Kourouklis *et al*. [15]. Nevertheless, contrary to their conclusions there is no formation of 3D Zn clusters at the beginning of the growth. We show that on both faces, there is formation of one flat dense (111) monolayer of Zn covering the entire substrate surface. From a detailed analysis of LEED and Moiré patterns observed by STM, we propose two ball models where the Zn monolayers are rotated with respect to the silver substrate, of about ± 1.5° on the Ag(111) face and of about ± 4.5° on the Ag(110) giving rise respectively to (√156x√156)R18° and c(12x6) superstructures. Hence Moiré patterns can also occur when the deposited elements are smaller than the substrate ones.

The analysis of Moiré patterns has proven to be a powerful technique to the determination of the atomic structure of a monolayer. Such method has been used for the determination of silicene structure on silver [17,18], graphene on metal [19] or metal on metal [20,21].

**Acknowledgements**

The authors gratefully acknowledge G. Tréglia and J. Rothwell for many helpful discussions and for critically reviewing this manuscript.